\documentclass[aps,prc,twocolumn,showpacs,preprintnumbers,nofootinbib,float,superscriptaddress,longbibliography]{revtex4-2}
\usepackage{graphicx,fancybox}
\usepackage{amsmath,amssymb}
\usepackage[colorlinks=true, pdfstartview=FitV, linkcolor=red, citecolor=blue, urlcolor=blue]{hyperref}
\usepackage{color}
\usepackage{soul}
\usepackage{array}
\usepackage{url}
\usepackage[utf8]{inputenc}
\usepackage{lineno}
\newcommand*\patchAmsMathEnvironmentForLineno[1]{%
\expandafter\let\csname old#1\expandafter\endcsname\csname #1\endcsname
\expandafter\let\csname oldend#1\expandafter\endcsname\csname end#1\endcsname
\renewenvironment{#1}%
{\linenomath\csname old#1\endcsname}%
{\csname oldend#1\endcsname\endlinenomath}}% 
\newcommand*\patchBothAmsMathEnvironmentsForLineno[1]{%
\patchAmsMathEnvironmentForLineno{#1}%
\patchAmsMathEnvironmentForLineno{#1*}}%
\AtBeginDocument{%
\patchBothAmsMathEnvironmentsForLineno{equation}%
\patchBothAmsMathEnvironmentsForLineno{align}%
\patchBothAmsMathEnvironmentsForLineno{flalign}%
\patchBothAmsMathEnvironmentsForLineno{alignat}%
\patchBothAmsMathEnvironmentsForLineno{gather}%
\patchBothAmsMathEnvironmentsForLineno{multline}%
}
\usepackage[normalem]{ulem}

\newcommand{\trento}{\textsc{trento}}
\newcommand{\pythia}{\textsc{pythia}}
\newcommand{\jetscape}{\textsc{jetscape}}
\newcommand{\matter}{\textsc{matter}}
\newcommand{\lbt}{\textsc{lbt}}
\newcommand{\colorlesshad}{\textsc{colorless hadronization}}
\newcommand{\pythiagun}{\textsc{pythiagun}}

\begin{document}

%%%%%% TITLE %%%%%%%%%%%%%%%
\title{Hard Photon Triggered Jets in $p$-$p$ and $A$-$A$ Collisions}

%%%%%% AUTHORS %%%%%%%%%%%%%
%%% JETSCAPE Author List

\author{C.~Sirimanna}
\email[Corresponding author: ]{chathuranga.sirimanna@duke.edu}
\affiliation{Department of Physics, Duke University, Durham, NC 27708.}
\affiliation{Department of Physics and Astronomy, Wayne State University, Detroit MI 48201.}

\author{Y.~Tachibana}
\affiliation{Akita International University, Yuwa, Akita-city 010-1292, Japan.}

\author{A.~Majumder}
\affiliation{Department of Physics and Astronomy, Wayne State University, Detroit MI 48201.}

%%%%%%%%%%%%%%%%%%%%%%%%%%%%%%%%
% %%%%%%%%%%%%%%%%%%%%%%%%%%%%%%%%
%%%%%% AUTHORS %%%%%%%%%%%%%
%%% JETSCAPE Author List

\author{A.~Angerami}
\affiliation{Lawrence Livermore National Laboratory, Livermore CA 94550.}

\author{R.~Arora}
\affiliation{Department of Computer Science, Wayne State University, Detroit MI 48202.}

\author{S.~A.~Bass}
\affiliation{Department of Physics, Duke University, Durham, NC 27708.}

\author{Y.~Chen}
\affiliation{Laboratory for Nuclear Science, Massachusetts Institute of Technology, Cambridge MA 02139.}
\affiliation{Department of Physics, Massachusetts Institute of Technology, Cambridge MA 02139.}
\affiliation{Department of Physics and Astronomy, Vanderbilt University, Nashville TN 37235.}

\author{R.~Datta}
\affiliation{Department of Physics and Astronomy, Wayne State University, Detroit MI 48201.}

\author{L.~Du}
\affiliation{Department of Physics, McGill University, Montr\'{e}al QC H3A\,2T8, Canada.}
\affiliation{Department of Physics, University of California, Berkeley CA 94270.}
\affiliation{Nuclear Science Division, Lawrence Berkeley National Laboratory, Berkeley CA 94270.}

\author{R.~Ehlers}
\affiliation{Department of Physics, University of California, Berkeley CA 94270.}
\affiliation{Nuclear Science Division, Lawrence Berkeley National Laboratory, Berkeley CA 94270.}

\author{H.~Elfner}
\affiliation{GSI Helmholtzzentrum f\"{u}r Schwerionenforschung, 64291 Darmstadt, Germany.}
\affiliation{Institute for Theoretical Physics, Goethe University, 60438 Frankfurt am Main, Germany.}
\affiliation{Frankfurt Institute for Advanced Studies, 60438 Frankfurt am Main, Germany.}

\author{R.~J.~Fries}
\affiliation{Cyclotron Institute, Texas A\&M University, College Station TX 77843.}
\affiliation{Department of Physics and Astronomy, Texas A\&M University, College Station TX 77843.}

\author{C.~Gale}
\affiliation{Department of Physics, McGill University, Montr\'{e}al QC H3A\,2T8, Canada.}

\author{Y.~He}
\affiliation{School of Physics and Optoelectronics, South China University of Technology, Guangzhou 510640, China}

\author{B.~V.~Jacak}
\affiliation{Department of Physics, University of California, Berkeley CA 94270.}
\affiliation{Nuclear Science Division, Lawrence Berkeley National Laboratory, Berkeley CA 94270.}

\author{P.~M.~Jacobs}
\affiliation{Department of Physics, University of California, Berkeley CA 94270.}
\affiliation{Nuclear Science Division, Lawrence Berkeley National Laboratory, Berkeley CA 94270.}

\author{S.~Jeon}
\affiliation{Department of Physics, McGill University, Montr\'{e}al QC H3A\,2T8, Canada.}

\author{Y.~Ji}
\affiliation{Department of Statistical Science, Duke University, Durham NC 27708.}

\author{F.~Jonas}
\affiliation{Department of Physics, University of California, Berkeley CA 94270.}
\affiliation{Nuclear Science Division, Lawrence Berkeley National Laboratory, Berkeley CA 94270.}

\author{L.~Kasper}
\affiliation{Department of Physics and Astronomy, Vanderbilt University, Nashville TN 37235.}

\author{M.~Kordell~II}
\affiliation{Cyclotron Institute, Texas A\&M University, College Station TX 77843.}
\affiliation{Department of Physics and Astronomy, Texas A\&M University, College Station TX 77843.}

\author{A.~Kumar}
\affiliation{Department of Physics, University of Regina, Regina, SK S4S 0A2, Canada.}
\affiliation{Department of Physics, McGill University, Montr\'{e}al QC H3A\,2T8, Canada.}

\author{R.~Kunnawalkam-Elayavalli}
\affiliation{Department of Physics and Astronomy, Vanderbilt University, Nashville TN 37235.}

\author{J.~Latessa}
\affiliation{Department of Computer Science, Wayne State University, Detroit MI 48202.}

\author{Y.-J.~Lee}
\affiliation{Laboratory for Nuclear Science, Massachusetts Institute of Technology, Cambridge MA 02139.}
\affiliation{Department of Physics, Massachusetts Institute of Technology, Cambridge MA 02139.}

\author{R.~Lemmon}
\affiliation{Daresbury Laboratory, Daresbury, Warrington, Cheshire, WA44AD, United Kingdom.}

\author{M.~Luzum}
\affiliation{Instituto  de  F\`{i}sica,  Universidade  de  S\~{a}o  Paulo,  C.P.  66318,  05315-970  S\~{a}o  Paulo,  SP,  Brazil. }

\author{S.~Mak}
\affiliation{Department of Statistical Science, Duke University, Durham NC 27708.}

\author{A.~Mankolli}
\affiliation{Department of Physics and Astronomy, Vanderbilt University, Nashville TN 37235.}

\author{C.~Martin}
\affiliation{Department of Physics and Astronomy, University of Tennessee, Knoxville TN 37996.}

\author{H.~Mehryar}
\affiliation{Department of Computer Science, Wayne State University, Detroit MI 48202.}

\author{T.~Mengel}
\affiliation{Department of Physics and Astronomy, University of Tennessee, Knoxville TN 37996.}

\author{C.~Nattrass}
\affiliation{Department of Physics and Astronomy, University of Tennessee, Knoxville TN 37996.}

\author{J.~Norman}
\affiliation{Oliver Lodge Laboratory, University of Liverpool, Liverpool, United Kingdom.}

\author{C.~Parker}
\affiliation{Cyclotron Institute, Texas A\&M University, College Station TX 77843.}
\affiliation{Department of Physics and Astronomy, Texas A\&M University, College Station TX 77843.}

\author{J.-F.~Paquet}
\affiliation{Department of Physics and Astronomy, Vanderbilt University, Nashville TN 37235.}

\author{J.~H.~Putschke}
\affiliation{Department of Physics and Astronomy, Wayne State University, Detroit MI 48201.}

\author{H.~Roch}
\affiliation{Department of Physics and Astronomy, Wayne State University, Detroit MI 48201.}

\author{G.~Roland}
\affiliation{Laboratory for Nuclear Science, Massachusetts Institute of Technology, Cambridge MA 02139.}
\affiliation{Department of Physics, Massachusetts Institute of Technology, Cambridge MA 02139.}

\author{B.~Schenke}
\affiliation{Physics Department, Brookhaven National Laboratory, Upton NY 11973.}

\author{L.~Schwiebert}
\affiliation{Department of Computer Science, Wayne State University, Detroit MI 48202.}

\author{A.~Sengupta}
\affiliation{Cyclotron Institute, Texas A\&M University, College Station TX 77843.}
\affiliation{Department of Physics and Astronomy, Texas A\&M University, College Station TX 77843.}

\author{C.~Shen}
\affiliation{Department of Physics and Astronomy, Wayne State University, Detroit MI 48201.}
\affiliation{RIKEN BNL Research Center, Brookhaven National Laboratory, Upton NY 11973.}

\author{M.~Singh}
\affiliation{Department of Physics and Astronomy, Vanderbilt University, Nashville TN 37235.}

\author{D.~Soeder}
\affiliation{Department of Physics, Duke University, Durham, NC 27708.}

\author{R.~A.~Soltz}
\affiliation{Department of Physics and Astronomy, Wayne State University, Detroit MI 48201.}
\affiliation{Lawrence Livermore National Laboratory, Livermore CA 94550.}

\author{I.~Soudi}
\affiliation{Department of Physics and Astronomy, Wayne State University, Detroit MI 48201.}
\affiliation{University of Jyväskylä, Department of Physics, P.O. Box 35, FI-40014 University of Jyväskylä, Finland.}
\affiliation{Helsinki Institute of Physics, P.O. Box 64, FI-00014 University of Helsinki, Finland.}

\author{J.~Velkovska}
\affiliation{Department of Physics and Astronomy, Vanderbilt University, Nashville TN 37235.}

\author{G.~Vujanovic}
\affiliation{Department of Physics, University of Regina, Regina, SK S4S 0A2, Canada.}

\author{X.-N.~Wang}
\affiliation{Key Laboratory of Quark and Lepton Physics (MOE) and Institute of Particle Physics, Central China Normal University, Wuhan 430079, China.}
\affiliation{Department of Physics, University of California, Berkeley CA 94270.}
\affiliation{Nuclear Science Division, Lawrence Berkeley National Laboratory, Berkeley CA 94270.}

\author{X.~Wu}
\affiliation{Department of Physics, McGill University, Montr\'{e}al QC H3A\,2T8, Canada.}
\affiliation{Department of Physics and Astronomy, Wayne State University, Detroit MI 48201.}

\author{W.~Zhao}
\affiliation{Department of Physics and Astronomy, Wayne State University, Detroit MI 48201.}
\affiliation{Department of Physics, University of California, Berkeley CA 94270.}
\affiliation{Nuclear Science Division, Lawrence Berkeley National Laboratory, Berkeley CA 94270.}

\collaboration{JETSCAPE Collaboration}

%%%%%% ABSTRACT %%%%%%%%%%%%
\begin{abstract}
An investigation of high transverse momentum (high-$p_T$) photon triggered jets in proton-proton ($p$-$p$) and ion-ion ($A$-$A$) collisions at $\sqrt{s_{NN}} = 0.2$ and $5.02~\mathrm{TeV}$ is carried out, using the multistage description of in-medium jet evolution. 
Monte Carlo simulations of hard scattering and energy loss in heavy-ion collisions are performed using parameters tuned in a previous study of the nuclear modification factor ($R_{AA}$) for inclusive jets and high-$p_T$ hadrons.
We obtain a good reproduction of the experimental data for photon triggered jet $R_{AA}$, as measured by the ATLAS detector, the distribution of the ratio of jet to photon $p_T$  ($X_{\rm J \gamma}$), measured by both CMS and ATLAS, and the photon-jet azimuthal correlation as measured by CMS.
We obtain a moderate description of the photon triggered jet $I_{AA}$, as measured by STAR. 
A noticeable improvement in the comparison is observed when one goes beyond prompt photons and includes bremsstrahlung and decay photons, revealing their significance in certain kinematic regions, particularly at $X_{J\gamma} > 1$. 
Moreover, azimuthal angle correlations demonstrate a notable impact of non-prompt photons on the distribution, emphasizing their role in accurately describing experimental results.
This work highlights the success of the multistage model of jet modification to straightforwardly predict (this set of) photon triggered jet observables. This comparison, along with the role played by non-prompt photons, has important consequences on the inclusion of such observables in a future Bayesian analysis. 
\end{abstract}

%%%%%% ABSTRACT %%%%%%%%%%%%
\maketitle

%%%%%%%%%%%%%%%%%%%%%%%%%%%%
%%%%%% MAIN TEXT %%%%%%%%%%%
%%%%%%%%%%%%%%%%%%%%%%%%%%%%

%%%%%% INTRODUCTION %%%%%%%%
\section{Introduction}
\label{Section:Intro}

The modification of inclusive hard jets, as they propagate through the quark-gluon plasma (QGP) produced in a heavy-ion collision, at the Relativistic Heavy-ion Collider (RHIC) or the Large Hadron Collider (LHC)~\cite{Song:2010mg,Bernhard:2019bmu,ALICE:2022wpn}, has now been measured over a wide range of jet transverse momenta ($p_T$), jet radius ($R$), centrality and energy of collision~\cite{ATLAS:2018gwx,ALICE:2019qyj,CMS:2016svx,CMS:2016uxf,STAR:2020xiv,CMS:2021vui}. A variety of theoretical approaches have been proposed to describe the modification of these jets in the QGP~\cite{Majumder:2009ge,Sirimanna:2021sqx,Arnold:2002ja,Gyulassy:2000fs,Burke:2013yra}. Several of these approaches, which typically focus on the energy lost by a single parton, in different epochs of jet evolution, in a dense medium~\cite{Majumder:2010qh}, have been implemented within Monte-Carlo jet shower routines~\cite{Majumder:2013re,Cao:2017qpx,He:2015pra,Luo:2023nsi,Schenke:2009gb,Shi:2022rja}.

Given extensive data and a multitude of observables, event generation approaches, 
where different observables can be built from the same set of 
generated events~\cite{Cao:2024pxc}, are clearly superior to single observable 
focused calculations, in the sense that they allow for multiple constraints on the same calculation from different data sets. 
Current event generators simulate almost every aspect of the collision. 
The bulk sector is simulated using a multistage approach including an initial stage of the two incoming nuclei~\cite{Moreland:2014oya,Shen:2017bsr,Schenke:2012wb}, a pre-equilibrium stage~\cite{Kurkela:2018vqr,Vredevoogd:2008id}, a locally thermalized viscous fluiddynamical stage~\cite{Shen:2014vra,Gale:2013da}, followed by hadronization and an interacting expanding hadron cascade~\cite{SMASH:2016zqf}. 
Comparing a sophisticated multistage generator (with several parameters) to experimental data requires an equally sophisticated machine learning framework. 
At this time, leading approaches for such a framework rely on Bayesian statistics~\cite{Novak:2013bqa,JETSCAPE:2020shq,JETSCAPE:2020mzn}. 
State-of-the-art simulations of jets (or the hard sector in general), use the same initial state and space-time profile which results from the best fit (or posterior distribution) in comparison to bulk observables, to determine the location of the hard scattering and the medium through which the jets propagate.

For some time, there remained the hope that the hard sector could be calculated using a straightforward energy loss model, or a single-stage generator, cast within the medium generated by the multistage bulk simulation. Different approximations, which applied to different stages of jet evolution, which led to differing energy loss models, were applied to the calculation of the suppression of the binary collision scaled yield of inclusive leading hadrons, measured using the nuclear modification factor, $R_{AA}$~\cite{PHENIX:2001hpc,Burke:2013yra}.

This first systematic comparison of different energy loss models (and single stage simulators), on the same bulk medium~\cite{Burke:2013yra}, focused only on leading hadron suppression in the most central events at $\sqrt{s_{NN}}=0.2$~TeV (RHIC) and 2.76~TeV (LHC), did yield somewhat consistent values of the jet transport coefficient $\hat{q}$, defined as the mean squared transverse momentum exchanged per unit length between a hard parton and the medium~\cite{Baier:2002tc,Majumder:2012ti,Kumar:2020wvb}.
However, even in this limited comparison, the fitted normalization of $\hat{q}$ at RHIC and LHC energies was different, with no calculation connecting the two values.  

Several attempts to simultaneously describe the spectra of inclusive jets and leading hadrons in heavy-ion collisions using a single stage energy loss model (with the same parameters), e.g. only \matter~\cite{Cao:2017qpx}, or only \lbt~\cite{Cao:2017hhk,He:2018xjv} etc., have not been successful. This led to the multistage approach, first proposed as a possible solution in Ref.~\cite{JETSCAPE:2017eso} and later theoretically justified in Ref.~\cite{Caucal:2018dla}. Realistic simulations using the multistage model have been quite successful in simultaneously describing the suppression of leading hadrons and jets (in all bins below 50\% centrality, at top RHIC, to all LHC energies)~\cite{JETSCAPE:2022jer}, suppression of heavy flavors (at LHC)~\cite{JETSCAPE:2022hcb}, and a wide variety of jet substructure measurements (also at LHC energies)~\cite{JETSCAPE:2023hqn}. 
In fact, recent work has demonstrated that in the absence of a multistage generator, it would not be possible to simultaneously describe jets and leading hadrons at any energy~\cite{Modarresi-Yazdi:2024vfh}.

Comparing an extensive sophisticated multistage simulator to a large set of experimental data involves two steps: First, one needs to demonstrate that the physics model \emph{can} describe the chosen data set; that varying parameters of the theory result in simulation results that can bracket the data points. 
In the second stage, a Bayesian analysis is performed by varying the entire parameter set within a prior range. 
This process generates predictions that fully encapsulate the selected data points. 
Using procedures such as machine learning and emulation, a cross-probability distribution of parameters is generated, resulting in a posterior distribution of predictions that is expected to closely align with the data points. 
This method allows one to (i) study the efficacy of the chosen model, (ii) explore cross correlations between different parameters, and (iii) isolate parameters that are actually constrained by the given data set. Parameters of a model typically represent or parametrize physical properties, e.g., $\hat{q}$, viscosities, thermalization times, etc. Bayesian analysis over an \emph{appropriately wide} data set is expected to yield realistic distributions of these physical properties.  

One then expands the data set and repeats the process, which again starts with a physics analysis to ``survey" the new data. 
Based on this survey, one may either run a more rigorous Bayesian analysis with the same model on the wider data set to obtain modifications of the parameter distributions and correlations, or, depending on the new data, introduce further extensions or improvements in the physical model. 
It is often the case that new data sets explore regimes outside the validity of a given approximation used to simplify a model or expedite its runtime, e.g., the lack of energy loss in the hadronic phase limits the applicability of the model of Refs.~\cite{JETSCAPE:2022jer,JETSCAPE:2022hcb,JETSCAPE:2023hqn} to centralities more peripheral than 50\%, which include a much larger bulk hadronic phase than a partonic phase. Physics surveys of new data sets are thus essential as they allow for an exploration of the new data set prior to a computationally involved Bayesian routine. 
Unlike the case in the typical Bayesian analysis, every comparison with data involves a simulation and not an emulation. The actual model is evaluated for every choice of parameters. 
They also afford a different fitting approach where theory uncertainties or biases can be approximately incorporated. 

The triple goals of the \jetscape\ collaboration are to construct an event generator framework within which one can build and improve event generators for heavy-ion collisions~\cite{Putschke:2019yrg}, to run wide-ranging physics analyses to compare these with experimental data to clearly delineate the range of validity of a given generator (physical model), to exhaustively compare each physical model with the largest data set to which the model applies via Bayesian analysis and extract the distribution and correlations between model parameters. 

The first physical model survey was published in Ref.~\cite{JETSCAPE:2017eso} and the corresponding Bayesian analysis restricted to only leading hadrons at top RHIC and one LHC energy (2.76~TeV) published in Ref.~\cite{JETSCAPE:2021ehl:Manual}. The model was then further developed to include coherence effects, with physics surveys published in Ref.~\cite{JETSCAPE:2022jer,JETSCAPE:2022hcb,JETSCAPE:2023hqn}, the corresponding Bayesian analysis which only includes leading hadrons and jets at all energies and centralities $\lessapprox 50\%$ has recently appeared in Ref.~\cite{JETSCAPE:2024cqe}. The current paper will attempt to answer the question of the next set of data that should be included in a future more extensive Bayesian analysis that may provide stronger constraints. 

Extending from the case of leading hadrons and inclusive jets, the next obvious set of data would involve jet substructure. A physics survey of a selection of jet substructure observables~\cite{JETSCAPE:2023hqn}, based on the parameters extracted from the comparison with leading hadrons~\cite{JETSCAPE:2022jer}, demonstrates a successful application of the model. As a result, including substructure observables should lead to an edification of the posterior probability distributions. Preliminary results indicate this to be the case~\cite{JETSCAPE:2024ofm}.  

To further probe the dynamics of energy loss, a natural extension of the data set may involve coincidence or triggered observables. 
One triggers on a specific particle or jet within a range of $p_T$, and then studies jets in the subset of events with that hard particle (jet). 
It has been proposed that jets recoiling off a hard photon allow for a more detailed study of energy loss of jets~\cite{Wang:1996yh}. 
The measured energy and transverse momentum of the photon introduce strong constraints on the energy of the recoiling jet. 
Unlike the case of a hadron trigger, triggering on a photon does not introduce any surface bias on the jet production point. 
As a result, one explores the energy loss of the jet in more or less the same medium as in the case of inclusive jet observables. 
The main difference is in the flavor of the jet; except in very rare cases, the recoiling hard parton that initiates the jet is a quark.

This physics survey of photon triggered jet observables will be split into two parts. In the current effort, we will focus on the entire jet recoiling off the photon. This will include observables such as the photon triggered jet $R_{AA}$, the transverse momentum imbalance $X_{J \gamma}$ and its centrality dependence, the azimuthal angular distribution, etc. Studies looking inside the recoiling jet, i.e., photon triggered jet substructure, will be included in a follow-up companion paper.

Extending the surveys of Refs.~\cite{JETSCAPE:2022jer,JETSCAPE:2022hcb,JETSCAPE:2023hqn}, to the case of photon triggered jets is not entirely straightforward. Hard photons with energies above 40~GeV are mostly produced in the initial hard scatterings. Given the energy of the jet, these may be produced lower down in the Multi-Particle Interaction (MPI) tower of hard scatterings. More importantly, it is possible to obtain hard and \emph{isolated} photons from the initial hard bremsstrahlung off a hard quark. 

In this paper, we will demonstrate that the primary choice of parameters extracted in Refs.~\cite{JETSCAPE:2022jer,JETSCAPE:2022hcb,JETSCAPE:2023hqn} is sufficient to explain a majority of the photon triggered jet results, if one includes non-prompt photons, i.e., bremsstrahlung photons. As will be shown below, simulating bremsstrahlung processes in addition to prompt photon production will require compute intensive simulations, as in these cases, one has to turn on both the electromagnetic (EM) and strong interaction (QCD) channels in typical simulators and collect sufficient statistics. In the case of purely prompt photon simulations, one can restrict the hard scattering to only events that produce at least one hard photon. However, to include bremsstrahlung photons, one turns on full QCD and EM cross sections and waits for a rare photon to be produced. 
In the calculations presented below, we will include results from both types of simulations: purely prompt photons, and simulations including prompt and bremsstrahlung photons.

%% Paper Organization
The paper is organized as follows. In Sec.~\ref{Section:model}, we recapitulate the salient features of the \matter\ + \lbt\ model of jet modification used within the \jetscape\ framework. In Sec.~\ref{Section:analysis}, we discuss issues related to the simulation of isolation cuts, jet reconstruction, and the smearing function used by the CMS experiment. Results of our comparisons with experimental data from ATLAS, CMS, and STAR are presented in Sec.~\ref{Section:Results}. An outlook is presented in Sec.~\ref{Section:Results}.

%%%%%% Model %%%%%%%%%%%
\section{Model}
\label{Section:model}
The \jetscape\ framework is a software framework that allows users to develop event generators for $p$-$p$ and $A$-$A$ collisions.
In this study, we utilize the publicly available \jetscape\ code package, which provides 
%a flexible framework for Monte Carlo event generation with 
default modules describing the specific physical processes involved in heavy-ion collisions. 
In addition to the ability to freely combine these default modules, users can test their own models by developing the requisite software in the simple \jetscape\ module format, which enables the replacement of default modules. 
One may replace an arbitrary number of modules, and keep the remainder as is. 
Such an approach eliminates the need for any one user to develop the extensive models typically required to realistically simulate heavy-ion collisions.

In this paper, we present the results of simulations using a configuration referred to as JETSCAPEv3.5 AA22 tune~\cite{JETSCAPE:2022jer, JETSCAPE:2022hcb, JETSCAPE:2023hqn}, which is built exclusively with the default \jetscape\ modules. 
All parameters related to this tune were chosen to fit only the single inclusive high-$p_{T}$ particle and jet $R_{AA}$s in Ref.~\cite{JETSCAPE:2022jer}, and were not retuned for any other observables, including the photon triggered jet observables presented in this paper.

\subsection{Framework Setup}
\label{Section:framework}
In the configuration of JETSCAPEv3.5 AA22 tune, the in-medium parton shower evolution of jets is managed by the multistage \trento+\pythia+\matter+\lbt\ setup. 
The location of the hard scattering is determined by the same initial state profile used to initiate the bulk evolution~(\trento). 
The initial state radiation (ISR) from the incoming hard partons and multiple hard scatterings (MPI) are simulated using the \pythia\ generator, with final state radiation (FSR) turned off.
Hard jet partons created in the hard scattering process are first sent to the \matter\ module~\cite{Majumder:2013re,Cao:2017qpx}. 

In \matter, the evolution of partons with large virtuality is described by the vacuum-like virtuality-ordered shower development, accompanied by the radiation of partons \emph{and photons}.
For parton radiation within a dense medium environment, suppressed medium effects are incorporated. 
The medium effect is estimated by incorporating modified coherence effects~\cite{Kumar:2019uvu} that suppress jet-medium interactions as the virtuality increases within the higher twist formulation~\cite{Abir:2014sxa,Abir:2015hta}. 
Unlike FSR in \pythia, \matter\ keeps track of the light-cone location of the partons as they shower, thereby providing both the light-cone momentum and location (which are not quantum conjugates) to the next energy loss module.

The description by the virtuality-ordered splitting becomes inapplicable as virtuality approaches the accumulation of transverse momentum acquired from the medium within the formation time. 
Thus, partons with reduced virtuality due to shower evolution in \matter\, transition to the transport-theory-based description provided by the \lbt\, module~\cite{Wang:2013cia,He:2015pra,Cao:2016gvr}, relying on the on-shell approximation. 
In the \lbt\ phase, any photon radiation is currently not implemented.
The assumption is that, at lower virtualities, any radiated photons would be collinear with the jet, most likely appearing within the jet cone, and thereby failing the isolation cut criteria, discussed below. 

The switching between the \matter\ and \lbt\ modules, utilizing the multistage description functionality of \jetscape, is performed bidirectionally on a parton-by-parton basis, using a switching virtuality parameter $Q_{\mathrm{sw}}$.
Inside the QGP medium with a temperature above $T_{c} = 0.16$ GeV, \matter\ is assigned for partons with virtuality above $Q_{\mathrm{sw}}$, while \lbt\ is assigned for partons with virtuality below $Q_{\mathrm{sw}}$. 
In this tune, switching virtuality is set as $Q_{\mathrm{sw}}=2$~GeV. 
After escaping from the QGP medium, partons undergo vacuum shower evolution by \matter, with the medium effect turned off until their virtualities descend to the cut-off scale $Q_{\mathrm{min}}^2=1\,{\mathrm{GeV}}^2$.

%Hadronization
These partons, upon completing their shower evolution, are directed to the \colorlesshad\ module, where they undergo hadronization based on the Lund string model of \pythia\ 8. 
As will be discussed also later in Sec.~\ref{Section:analysis}, both in \matter\ and \lbt, energy-momentum deficits, commonly called \emph{hole} partons~\cite{Tachibana:2020mtb} or \emph{negative} partons~\cite{Luo:2023nsi}, occur in the medium along with recoil partons due to scatterings with the medium partons. 
While the jet shower partons are hadronized together with recoil partons, the hole partons are hadronized separately. 
The recoil-hole method is an approximate pQCD based means to keep track of the excitation of the medium surrounding a jet, due to its traversal. 
This method is known to produce a less than accurate representation of the jet wake at larger jet radii ($R\gtrsim 0.5$) or angles away from the jet~\cite{Tachibana:2017syd,Yang:2022nei}.

%Medium
To calculate the medium effects on a jet parton, one needs to use information on the local medium at the parton location. 
For this purpose, the spacetime profile of the medium and the initial parton production location are required. 
For the background medium profile, we use pre-generated soft events from Bayesian-calibrated calculations~\cite{Bernhard:2019bmu} of event-by-event $(2+1)$-dimensional [$(2+1)$-D] free-streaming pre-equilibrium evolution~\cite{Liu:2015nwa}, followed by viscous hydrodynamic evolution by $(2+1)$-D \textsc{vishnu}~\cite{Shen:2014vra} with the \trento~\cite{Moreland:2014oya} initial conditions. 
The position of the partons produced in the initial hard process is obtained by sampling the $N_{\mathrm{coll}}$ distribution in the transverse plane ($\eta_{s}=0$) for the $A$-$A$ collisions from the same \trento\ initial condition. 

% PP19 tune
For the baseline study of heavy-ion collisions, we conducted $p$+$p$ collision simulations using the \jetscape\ framework with the JETSCAPE PP19 tune, as detailed in Ref.~\cite{JETSCAPE:2019udz}, where the parton shower evolution is solely managed by \matter, without jet parton-medium interactions.

\subsection{Initial hard process}
\label{Section:hard}
To generate the initial hard partons, we employ the \pythiagun\ module, which invokes the functionalities of \pythia\ 8~\cite{Sjostrand:2019zhc} for the generation of initial hard processes, with initial state radiation (ISR) and multiparton interaction (MPI) enabled. 
Final state radiation (FSR) is disabled in the \pythiagun\ module of \jetscape\ by default, and all resultant partons are directly sent to the \matter\ module for virtuality-ordered shower evolution. 

In this study, to systematically explore contributions to photon triggered jet events from processes beyond prompt photon production, we compare two different setups for the initial hard scattering channels: 
\begin{enumerate}
\item Full Events: the set of events with initial hard scatterings in which all hard QCD 2-to-2 channels and prompt photon production channels are on, achieved by setting \texttt{HardQCD:all=on} and \texttt{PromptPhoton:all=on}. 

\item Prompt Photon Events: the set of events with initial hard scatterings involving only prompt photon production channels by setting \texttt{HardQCD:all=off} and \texttt{PromptPhoton:all=on}.
\end{enumerate}
Other than the hard scattering channels in the \pythiagun\ module, all configurations and parameters are the same in these two event sets. 
The difference between the photon triggered jet yields from these two event sets is solely due to the contribution of fragmentation photons from jets generated by QCD hard scatterings that do not produce prompt photons.

\subsection{Hadronic decays}
\label{Section:decay}

Both ATLAS and CMS applied data-driven purity corrections to account for hadronic decay contributions to direct photons. However, these methods differ between the experiments. In this study, we restrict $\pi^0$ decays by using \texttt{111:mayDecay = off} in the hadronization module, since $\pi^0$ decays are the primary source of photon production within all hadronic decays. 

With this constraint, we determined that the contribution of hadronic decay photons to isolated direct photons is minimal compared to other included sources of photons. Thus the inclusion of hadronic decays with the $\pi^0$ decay restriction introduces only a negligible modification to the observables.

%%%%%% Analysis %%%%%%%%%%%
\section{Analysis Details}
\label{Section:analysis}

In this section, we will discuss specific details of the procedures used to gather and 
analyze simulation results. These involve specific isolation criteria to single out non-fragmentation photons at LHC energies, which are slightly modified due to the presence of recoils and holes in the simulation. The subtraction of holes to obtain the actual jet energies is then outlined. Following this, we discuss the smearing procedure used by the CMS detector, which is also reproduced in our results. Finally,  we outline the two different angular separation criteria at LHC and RHIC energies. 

\subsection{Isolation requirement for photons}

A prompt photon, originating from the initial hard scattering process, typically presents with a transverse momentum close to that of the pair-produced parent parton of the jet, even in the presence of the QGP medium, since photons do not exhibit strong interaction. 
However, photon-jet pairs may also arise from other photon production processes, such as jet fragmentation photon. 
To reduce the contamination by fragmentation photons, the isolation requirement, in which only photons with minimal transverse energy emissions within a fixed radius surrounding the photon are selected, is commonly employed. 
Such isolation cuts do not remove large angle bremsstrahlung photons.

In the \matter$+$\lbt\ simulations of \jetscape, scatterings between jet partons and medium partons lead to the production of hole and recoil partons. 
In this study, we take into account the contributions from holes in the isolation requirement of the photon triggered jet analysis. 
Thus, the accumulated transverse energy within the radius $R_{\mathrm{iso}}$ around the isolated photon candidate is calculated as: 
\begin{align}
E^{\mathrm{iso}}_{T}
&= 
\sum_{\substack{i\in \mathrm{shower}\\ \Delta r_i < R_{\mathrm{iso}}}}E_{T,i} - 
\sum_{\substack{i\in \mathrm{holes}\\ \Delta r_i < R_{\mathrm{iso}}}}E_{T,i}
-E_{T}^{\gamma}.
\label{eq:Isolation}
\end{align}
Here, $E_{T}^{\gamma}$ is the transverse energy and momentum of the isolated photon candidate. 
Both sums on the right-hand side are taken for particles with $\Delta r_i < R_{\mathrm{iso}}$, where $\Delta r_i = [(\eta_i - \eta_{\gamma})^2+(\phi_i - \phi_{\gamma})^2]^{1/2}$ 
is the radial distance from the isolated photon candidate. 
The first sum accounts for particles from the hadronization of jet shower partons, including the recoils, while the second sum is for the hadronized holes. 
If the candidate meets the condition $E^{\mathrm{iso}}_{T} < E^{\mathrm{iso,cut}}_{T}$ determined by a preset cut parameter $E^{\mathrm{iso,cut}}_{T}$, then it is triggered as an isolated photon.

\subsection{Jet reconstruction}
For events with a triggered isolated photon, jet reconstruction is performed to count photon triggered jets, taking into account the hole contribution. 
First, particles in the event other than hadronized holes are passed to the jet reconstruction by the anti-$k_{t}$ algorithm~\cite{Cacciari:2008gp}, with a jet cone size $R$ implemented in the \textsc{fastjet} package~\cite{Cacciari:2005hq, Cacciari:2011ma}. 
Then, the four-momenta of reconstructed jets are adjusted by subtracting the hole contribution: 
\begin{align}
p^\mu_{\mathrm{jet}}
&=p^\mu_{\mathrm{shower}}
-\sum_{\substack{i\in \mathrm{holes}\\ \Delta r_i < R}} p^\mu_i, 
\label{eq:neg_sub}
\end{align}
where $p^\mu_{\mathrm{shower}}$ is the four-momentum of jet reconstructed from particles other than those from holes.

\subsection{Smearing}
\label{subsec:smearing}
Both STAR and ATLAS results~\cite{ATLAS:2018dgb, ATLAS:2023iad} can be directly compared with results from Monte Carlo event generators since they used a two-dimensional unfolding method to correct their results for detector effects. However, for an accurate comparison with the CMS results~\cite{CMS:2017ehl}, a Gaussian smearing function needs to be applied. The transverse momentum of a jet ($p^{\mathrm{jet}}_{T}$), recoiling of a photon, is distributed away from its clustered value, using a narrow Gaussian function peaked at the actual $p^{\mathrm{jet}}_{T}$, with a standard deviation given as~\cite{CMS:2017ehl}, 
\begin{align}
\sigma \left( p^{\mathrm{jet}}_{T} \right) &= \sqrt{ C^2 + \frac{ S^2 }{ p^{\mathrm{jet}}_{T} } + \frac{ N^2 }{ \left( p^{\mathrm{jet}}_{T} \right)^2 }}.
\label{eq:sigma}
\end{align}

The width of the Gaussian narrows with increasing $p^{\mathrm{jet}}_{T}$. As a result, the smearing has the largest effect at lower values of $p^{\mathrm{jet}}_{T}$. In the subsequent section, we will find the largest effect of the smearing at lower $p^{\mathrm{jet}}_{T}$ in the CMS data.
The parameters are set to $C^2 = 0.0036$ and $S^2 = 1.5376$~GeV for both centralities of $0\%$--$10\%$ and $0\%$--$30\%$. Further, $N^2 = 70.8964$$~\mathrm{GeV}^2$ for $0\%$--$10\%$ and $N^2 = 46.6489$$~\mathrm{GeV}^2$ for $0\%$--$30\%$.

\subsection{Relative azimuth angle cut}
\label{subsec:azimuthCut}

For the photon triggered jet 
$R_{AA}$ and photon-jet transverse momentum imbalance 
($X_{J\gamma}$) distribution, jets are counted only when satisfying the relative azimuth angle cut condition: 
\begin{align}
\Delta \phi \equiv \lvert\phi_{\mathrm{jet}}-\phi_{\gamma}\rvert > \Delta\phi_{\mathrm{cut}} .
\end{align}
This ensures that the photon-jet pair is close to being back-to-back, increasing the contribution from the prompt photon production process.
The cut $\Delta\phi_{\mathrm{cut}}=7\pi/8$ is used by both the ATLAS and CMS detectors. 
Obviously, the relative azimuth angle cut is not applied for the photon-jet azimuthal angle ($\Delta \phi$) correlation. The STAR detector uses a cut of $\Delta\phi_{\mathrm{cut}}=3\pi/4$.

%%%%%% Results %%%%%%%%%%%%%
\section{Results}
\label{Section:Results}

As explained in the following subsections, photon triggered jet $R_{AA}$, photon-jet momentum imbalance ($\gamma$-jet asymmetry), and photon-jet correlation were investigated for $5.02$~TeV Pb-Pb collisions across different centralities using the \jetscape\ framework. These results were compared with the available CMS~\cite{CMS:2017ehl} and ATLAS~\cite{ATLAS:2018dgb, ATLAS:2023iad} results. The contribution from isolated non-prompt photons was studied by controlling the initial hard scattering processes as described in subsection \ref{Section:hard}. The combined prompt and non-prompt isolated photon triggered jet $I_{AA}$ was also calculated for Au-Au collisions at $\sqrt{s_{\rm NN}} = 0.2$~TeV and compared to data from STAR. We present and discuss each of these results in the subsections below.

\subsection{Photon triggered jet $R_{AA}$}
\label{Subsection:RAA}

\begin{figure*}[htb!]
\centering
\includegraphics[width=0.66\textwidth]{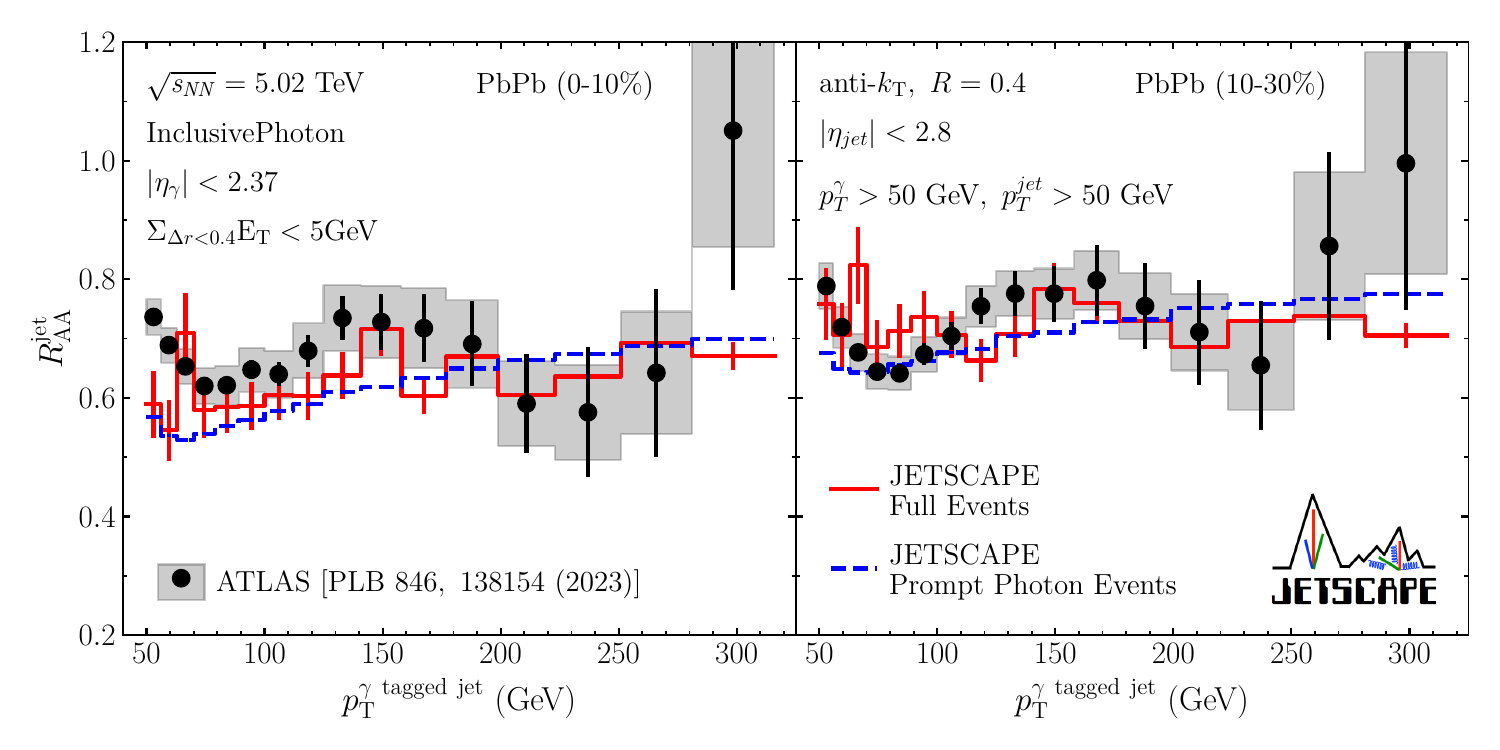}  
\caption{Nuclear modification factor $R_{AA}$ as a function of jet-$p_{T}$ for photon triggered jet for central $0\%$--$10\%$ (left) and semi-central $10\%$--$30\%$ (right) Pb-Pb collisions at $\sqrt{s_{NN}} = 5.02$~TeV. 
The results from \matter$+$\lbt\ within \jetscape\ for full events (solid lines) and prompt photon events (dashed lines) are compared with ATLAS data~\cite{ATLAS:2023iad}. }
\label{fig:RAA_ATLAS}
\end{figure*}

Figure~\ref{fig:RAA_ATLAS} shows the nuclear modification factor for photon triggered jets in Pb-Pb collisions at a $\sqrt{s_{NN}} = 5.02$~TeV. The \jetscape\ results are compared with recent ATLAS measurements~\cite{ATLAS:2023iad}. Isolated photons with $p_T^{\gamma} > 50$~GeV and $|\eta_{\gamma}| < 2.37$ were selected by setting $E_T^{\mathrm{iso}} \leq 5$~GeV and $R_{\mathrm{iso}} = 0.4$. Computationally, this is a rather demanding process as one has to weed through several simulations prior to events with the requisite photons. Although there is a blind pseudorapidity region $1.37 < |\eta_{\gamma}| < 1.52$ in the experimental analysis, we used the full pseudorapidity range in this analysis, as the effect from the blind region is insignificant. Reconstructed anti-$k_{t}$ jets, with a jet radius $R=0.4$, a $p_T^{\mathrm{jet}} > 50$~GeV, and an $|\eta_{\mathrm{jet}}| < 2.8$ were selected from events with the hard isolated photons. We designate these events as \emph{Full} events as pointed out in subsection~\ref{Section:hard}.
Jet populations were restricted by applying a cut on the relative angle with respect to the isolated photon $\Delta \phi_{\mathrm{cut}} > \frac{7}{8} \pi$.

Prompt photon production events are rare since the cross sections of those channels are small. Therefore, it is common practice to use modified event simulations that only contain prompt photon production channels. These events are identified as \emph{Prompt Photon} events, as explained in subsection~\ref{Section:hard}. 
Since these prompt photon events always contain a photon originating from the initial hard scattering, which is hard and more likely to satisfy the isolated condition, the statistical uncertainty is negligibly small. 
On the other hand, full events, which contain all hard QCD channels, have a significantly small number of isolated photon events. This introduces a relatively large statistical uncertainty. 
In both settings, the typical values in the experimental results are reproduced. 
The full event results capture the behavior of the experimental results slightly more closely, but with the current uncertainties, a definitive conclusion cannot yet be drawn. 

\subsection{Photon triggered jet $I_{AA}$}
\label{Subsection:IAA}

In this subsection, we present the results for the ratio for the photon triggered jet yields in central Au-Au and $p$-$p$ collisions:
\begin{align}
    I_{AA} = \frac{Y^{AA}(p_T^{\mathrm{jet}})}{Y^{pp}(p_T^{\mathrm{jet}})},
\end{align}
where $Y(p_T^{\mathrm{jet}}) = \frac{1}{N_{trig}} \int d\phi \frac{d^3 N_{jet}}{d p_T^{\mathrm{jet}} d\eta_{\mathrm{jet}} d\phi}$ is the differential jet yield per hard photon, in bins of $p_T$, $\eta$ and $\phi$. 
Jets with a jet cone size $R=0.2$ and $R=0.5$ are reconstructed using only charged particles with $|\eta| < 1.0$. A relative azimuthal angle cut, $\Delta \phi > 3 \pi / 4$ is applied to identify triggered jets as described in Sec.~\ref{subsec:azimuthCut}. 

\begin{figure*}[htb!]
\centering
\includegraphics[width=0.66\textwidth]{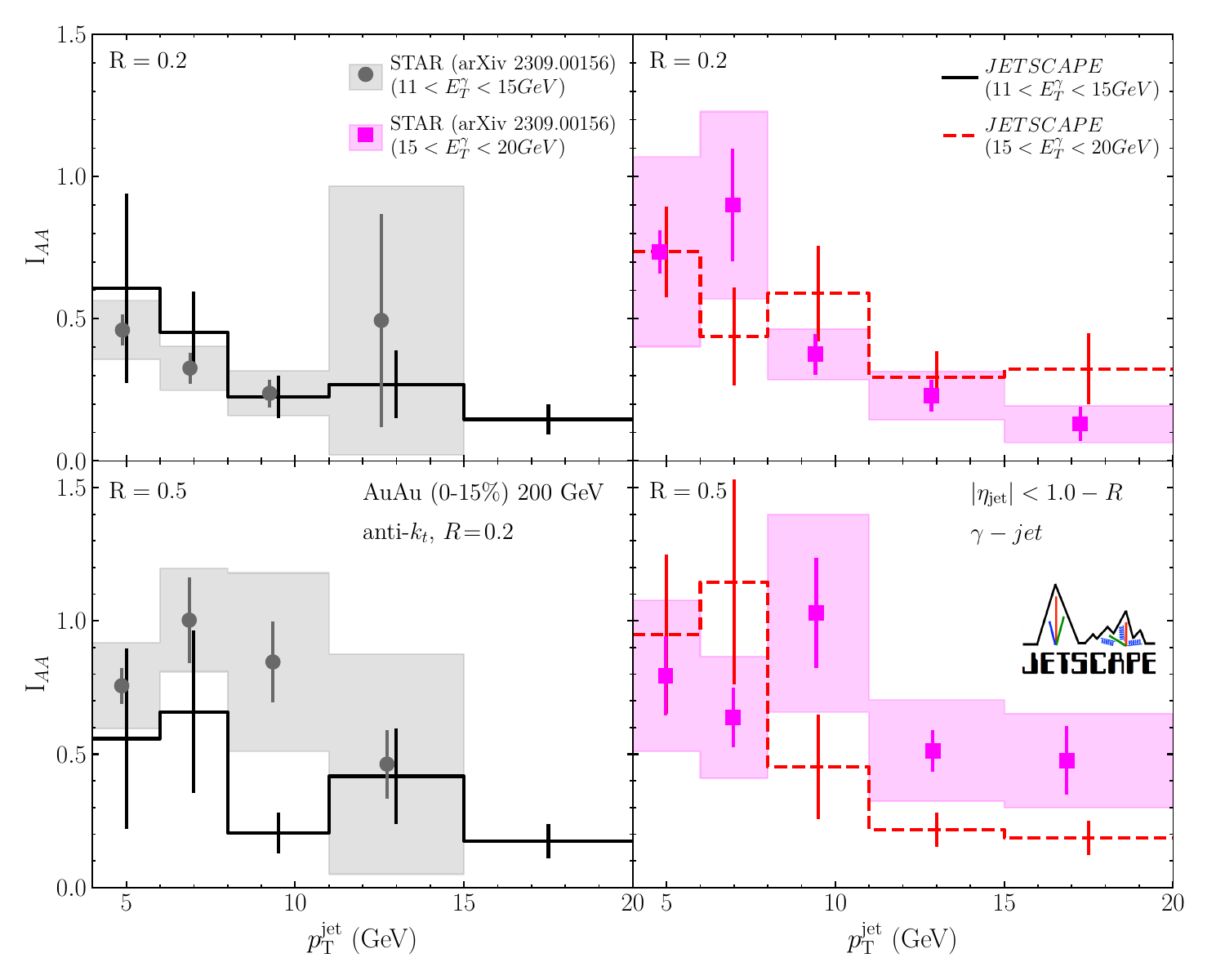}
\caption{Ratio of photon triggered jet yields in central $0\%$--$10\%$ Au-Au and $p$-$p$ collisions, $I_{AA}$ as a function of $p^{\mathrm{jet}}_{T}$ for two different jet radii, $R=0.2$ (left) and $R=0.5$ (right). The results from prompt and bremsstrahlung photon events generated using \matter$+$\lbt\ module combination within the \jetscape\ framework for $11 < E_T^\gamma < 15$~GeV (solid red lines) and $15< E_T^\gamma < 20$~GeV (dashed blue lines) are compared with STAR data \cite{STAR:2023pal,STAR:2023ksv}.}
\label{fig:IAA_STAR}
\end{figure*}

Figure~\ref{fig:IAA_STAR} compares the \jetscape\ results to STAR measurements of the $I_{AA}$ distribution for photon triggered jets in central Au-Au collisions at $\sqrt{s_{NN}} = 200~\text{GeV}$. In this measurement, direct photons are extracted from a distribution of the transverse shower profile and photon triggered jet distributions are obtained using a statistical method~\cite{STAR:2023pal,STAR:2023ksv}. In the simulation, prompt and bremsstrahlung photons without isolation are considered as direct photons in this comparison. Bremstrahlung photons are only produced in the MATTER portion, from partons with a large virtuality. As a result, these tend to be well separated from the majority of jet fragments. These direct photons are subject to a rapidity cut of $|\eta_\gamma| < 1.0$, while the triggered charged jets satisfy $|\eta_{\mathrm{jet}}| < 1.0 - R$.

Although the experimental data are accompanied by large systematic uncertainties, particularly at low  $p^{\mathrm{jet}}_{T}$, the \jetscape\ results show moderate agreement across all variations of the jet radius and $E_T^\gamma$ range. We remind the reader that there is no tuning of any parameter in these comparisons. The parameters were tuned in Ref.~\cite{JETSCAPE:2022jer}, and are used unchanged in this effort, for simulations at both RHIC and LHC energies.

\subsection{Photon-jet transverse momentum imbalance}
\label{Subsection:Asymmetry}
In this subsection, we present the results for the distribution of photon-jet transverse momentum imbalance: 
\begin{align}
\frac{1}{N_{\gamma}}\frac{dN_{\mathrm{jet}}}{dX_{J\gamma}},
\label{eq:x_dist}
\end{align}
where $N_{\gamma}$ is the number of triggered isolated photons, 
$N_{\mathrm{jet}}$ is the number of photon triggered jets, and $X_{J\gamma} = {p^{\mathrm{jet}}_{T}}/{p^{\gamma}_{T}}$ is the photon-jet transverse momentum imbalance. 
Here, the relative azimuthal angle cut $\Delta \phi > 7\pi/8$ is imposed.

\begin{figure*}[htb!]
\centering
\includegraphics[width=0.99\textwidth]{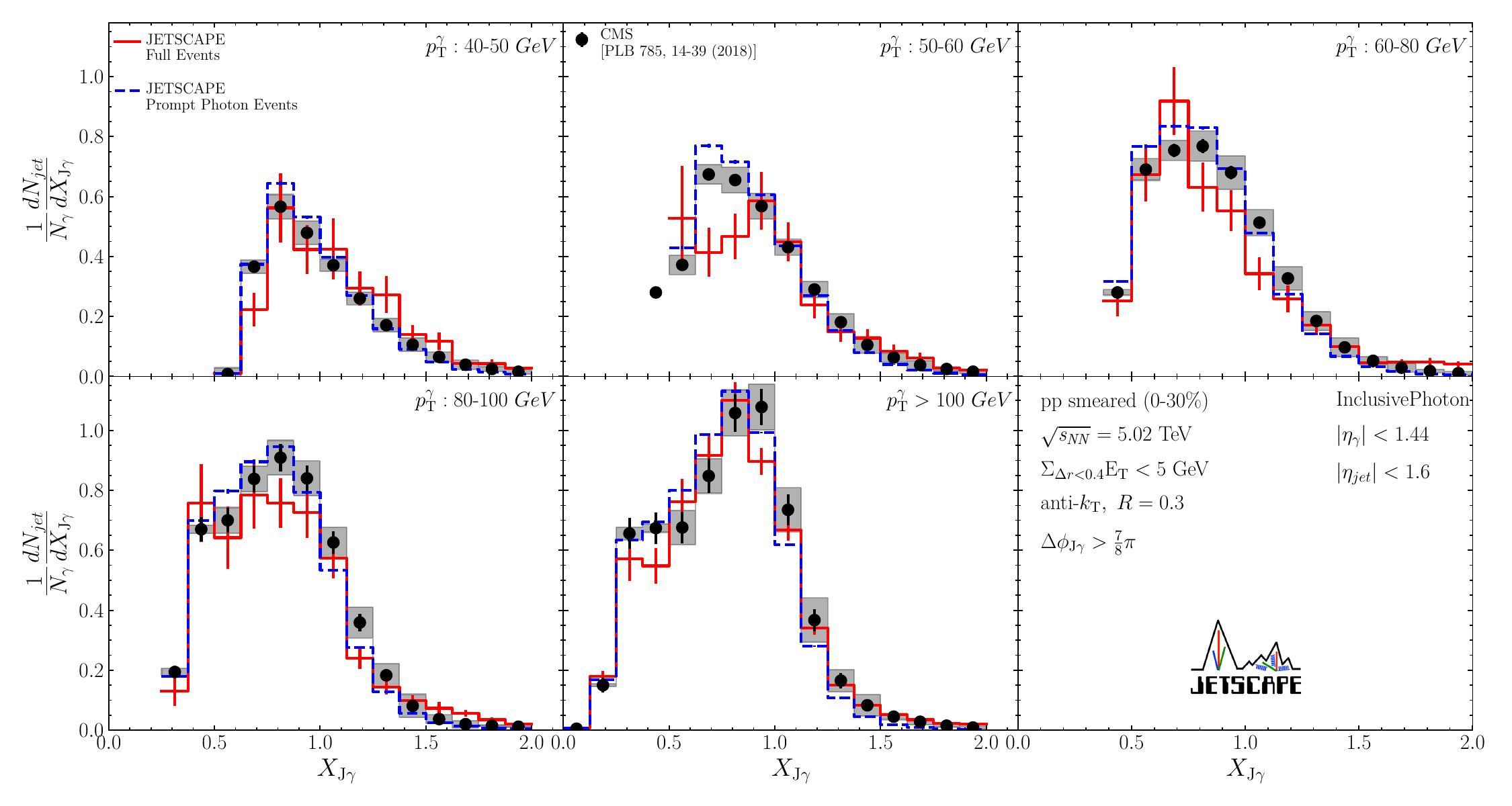}  
\caption{
Photon triggered jet transverse momentum imbalance in $p$-$p$ collisions at $\sqrt{s_{NN}} = 5.02$~TeV, smeared for $0$\%--$30$\%, for five different $p_T^{\gamma}$ intervals. 
The results from \jetscape\ for full events (solid lines) and prompt photon events (dashed lines) are compared with CMS data~\cite{CMS:2017ehl}. The peak in the distributions around $X_{\rm J \gamma} = p_T^{\rm Jet}/p_T^\gamma = 1$, corresponds to the condition where the $p_T$ of the photon is balanced with the recoiling jet. CMS data are not fully unfolded, and so an experimentally determined smearing of the $p_T$ distribution was applied to the simulation results prior to comparison with data (see text for details).}
\label{fig:Xj_CMS_pp_0_30}
\end{figure*}
\begin{figure*}[htb!]
\centering
\includegraphics[width=0.68\textwidth]{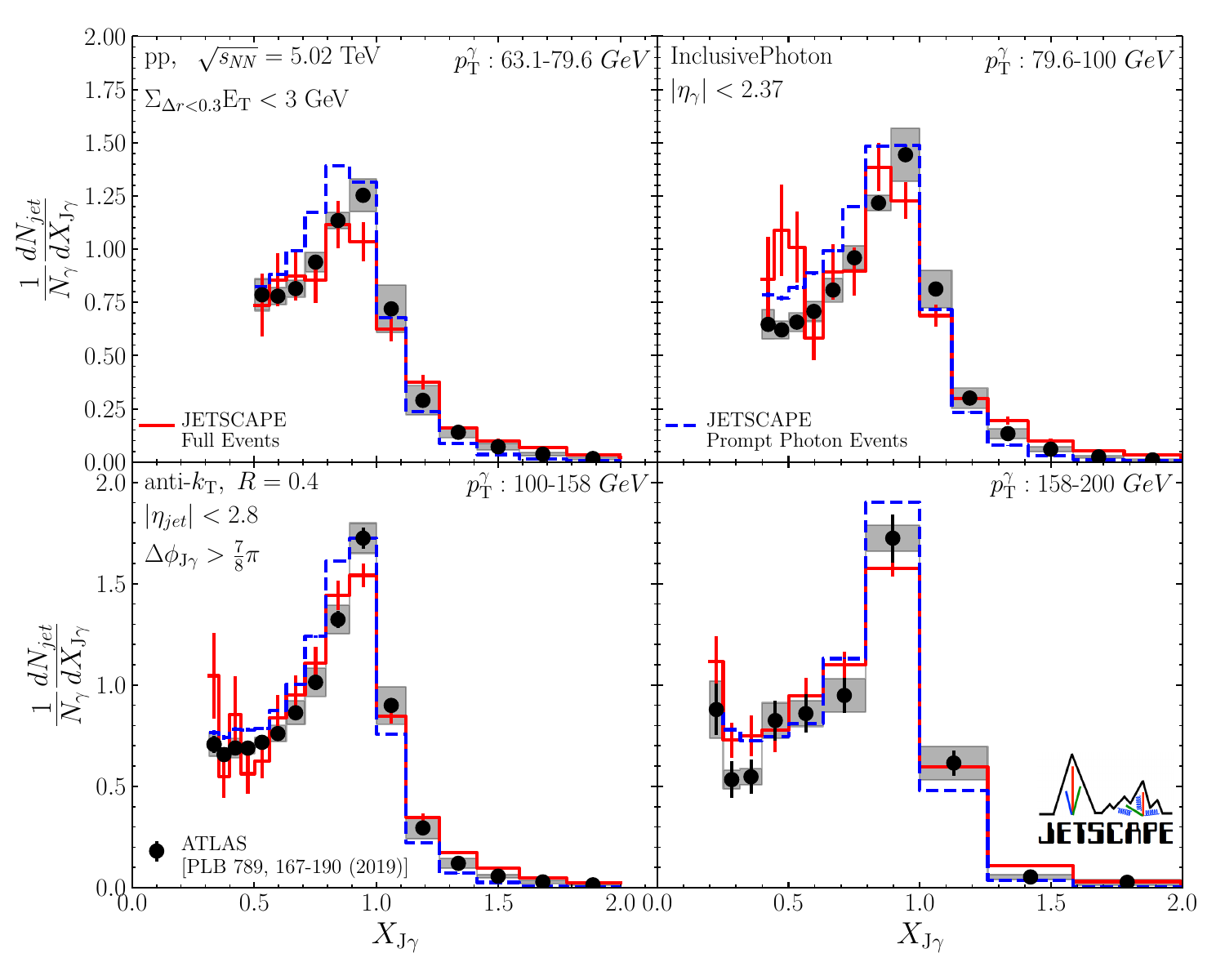} 
\caption{Photon triggered jet transverse momentum imbalance in $p$-$p$ collisions at $\sqrt{s_{NN}} = 5.02$~TeV for four different photon transverse momentum ($p_T^{\gamma}$) intervals ranging from $63.1$~GeV to $200$~GeV. 
The results from \jetscape\ for full events (solid lines) and prompt photon events (dashed lines) are compared with ATLAS data~\cite{ATLAS:2018dgb}.  The peak in all the distributions, at $X_{\rm J \gamma} = p_T^{\rm Jet}/p_T^\gamma = 1$, corresponds to the condition where the $p_T$ of the photon is balanced with the recoiling jet. ATLAS data are fully unfolded, and thus unlike the case in Fig.~\ref{fig:Xj_CMS_pp_0_30}, no smearing is applied to the simulation results.}
\label{fig:Xj_ATLAS_pp}
\end{figure*}

Figures~\ref{fig:Xj_CMS_pp_0_30} and \ref{fig:Xj_ATLAS_pp} show results for the $X_{J\gamma}$ distribution in $p$-$p$ collisions at $\sqrt{s_{NN}}=5.02$ TeV, compared to the CMS~\cite{CMS:2017ehl} and ATLAS~\cite{ATLAS:2018dgb} data, respectively. 
Since the ATLAS results were unfolded, our simulation results can be directly compared. 
In contrast, comparison with the CMS results requires applying the $p_T^{\mathrm{jet}}$ smearing described in \ref{subsec:smearing}. This smearing mechanism is consistently applied to all results compared with the CMS data in this study.
For the CMS results, isolated photons are selected with $|\eta_\gamma| < 1.44 $ and combined transverse energy $E_T^{\mathrm{iso}} < 5$ GeV within the cone of $R_{\mathrm{iso}}=0.4$ surrounding the photon. For the ATLAS results photons are isolated with $|\eta_\gamma| < 2.37 $ and $E_T^{\mathrm{iso}} < 3$ GeV within the cone of $R_{\mathrm{iso}}=0.3$ around the photon.

\begin{figure*}[htb!]
\centering
\includegraphics[width=0.99\textwidth]{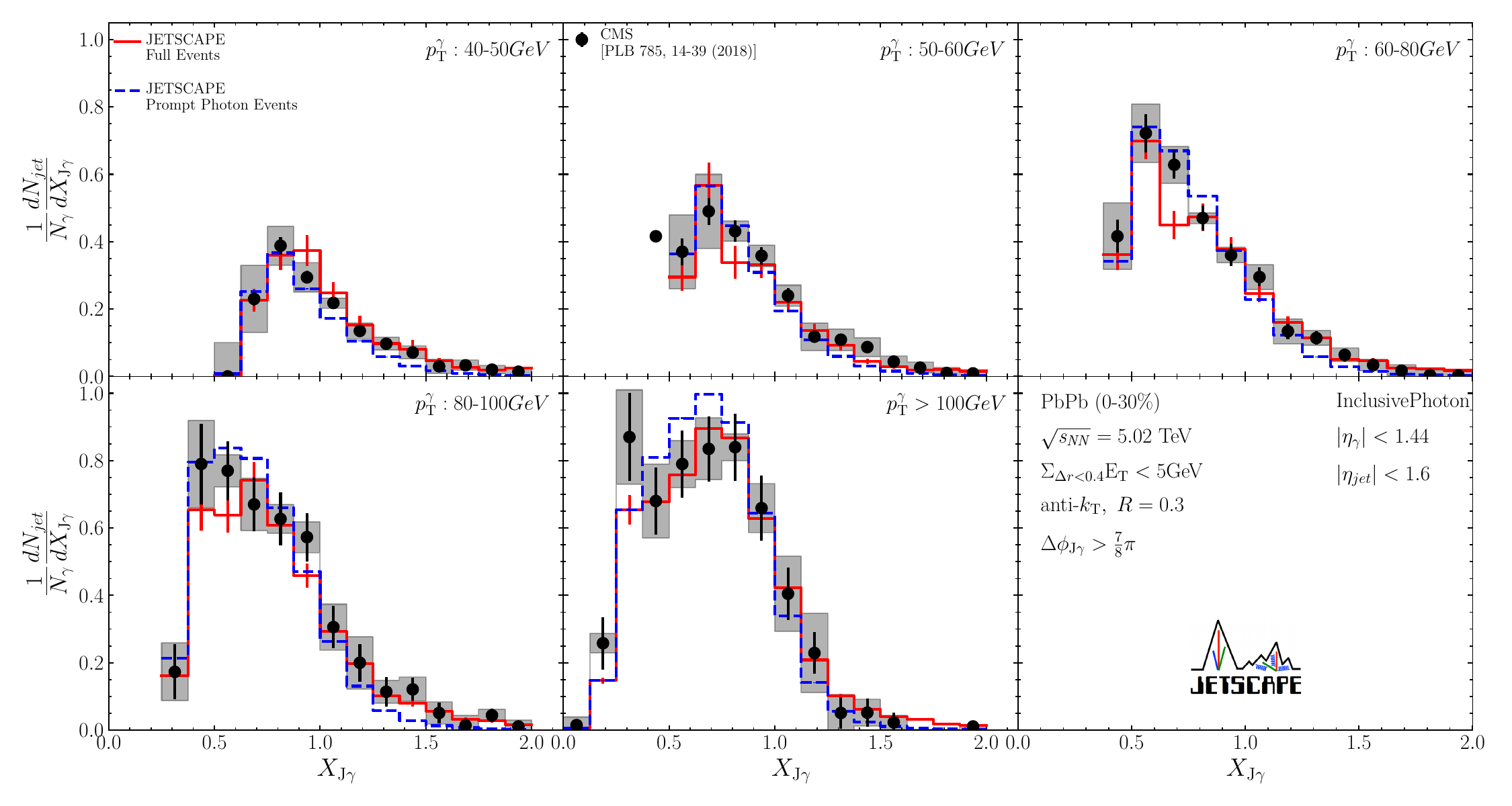}  
\caption{Photon triggered jet transverse momentum imbalance in Pb-Pb collisions of the $0\%$--$30\%$ centrality at $\sqrt{s_{NN}} = 5.02$~TeV for five $p_T^{\gamma}$ intervals. 
The results from \matter$+$\lbt\ within \jetscape\ for full events (solid lines) and prompt photon events (dashed lines) are compared with CMS data~\cite{CMS:2017ehl}. CMS data are not fully unfolded, and so an experimentally determined smearing of the $p_T$ distribution was applied to the simulation results prior to comparison with data (see text for details). }
\label{fig:Xj_CMS_PbPb_0_30}
\end{figure*}
\begin{figure*}[htb!]
\centering
\includegraphics[width=0.66\textwidth]{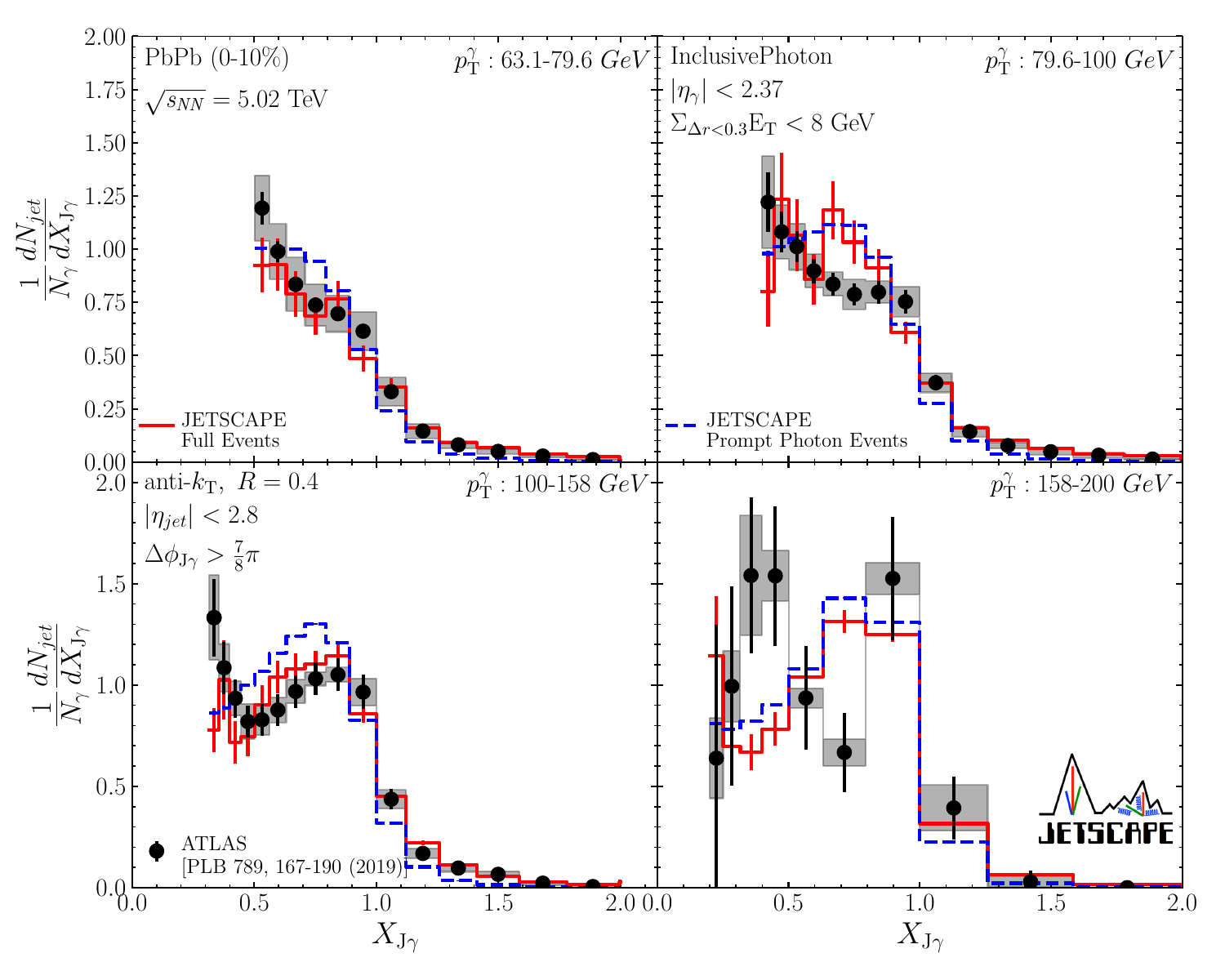}  
\caption{Photon triggered jet transverse momentum imbalance in Pb-Pb collisions of the $0\%$--$10\%$ centrality at $\sqrt{s_{NN}} = 5.02$~TeV for four different $p_T^{\gamma}$ intervals ranging from $63.1$~GeV to $200$~GeV. 
The results from \matter$+$\lbt\ within \jetscape\ for full events (solid lines) and prompt photon events (dashed lines) are compared with ATLAS data~\cite{ATLAS:2018dgb}. ATLAS data are fully unfolded, and thus unlike the case in Fig.~\ref{fig:Xj_CMS_PbPb_0_30}, no smearing is applied to the simulation results.}
\label{fig:Xj_ATLAS_0_10}
\end{figure*}
\begin{figure*}[htb!]
\centering
\includegraphics[width=0.66\textwidth]{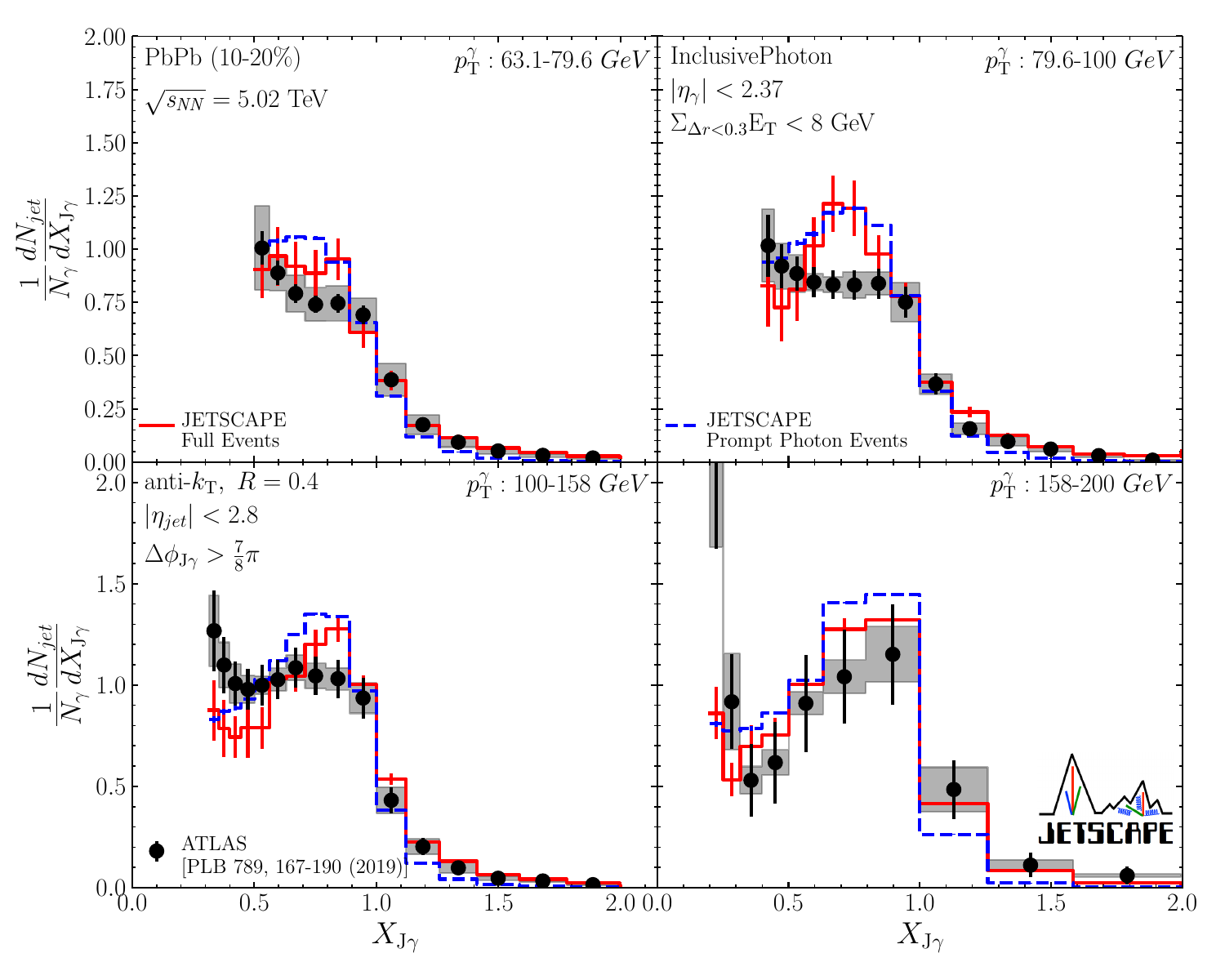}  
\caption{Same as Fig.~\ref{fig:Xj_ATLAS_0_10} for the $10\%$--$20\%$ centrality}
\label{fig:Xj_ATLAS_10_20}
\end{figure*}
\begin{figure*}[htb!]
\centering
\includegraphics[width=0.66\textwidth]{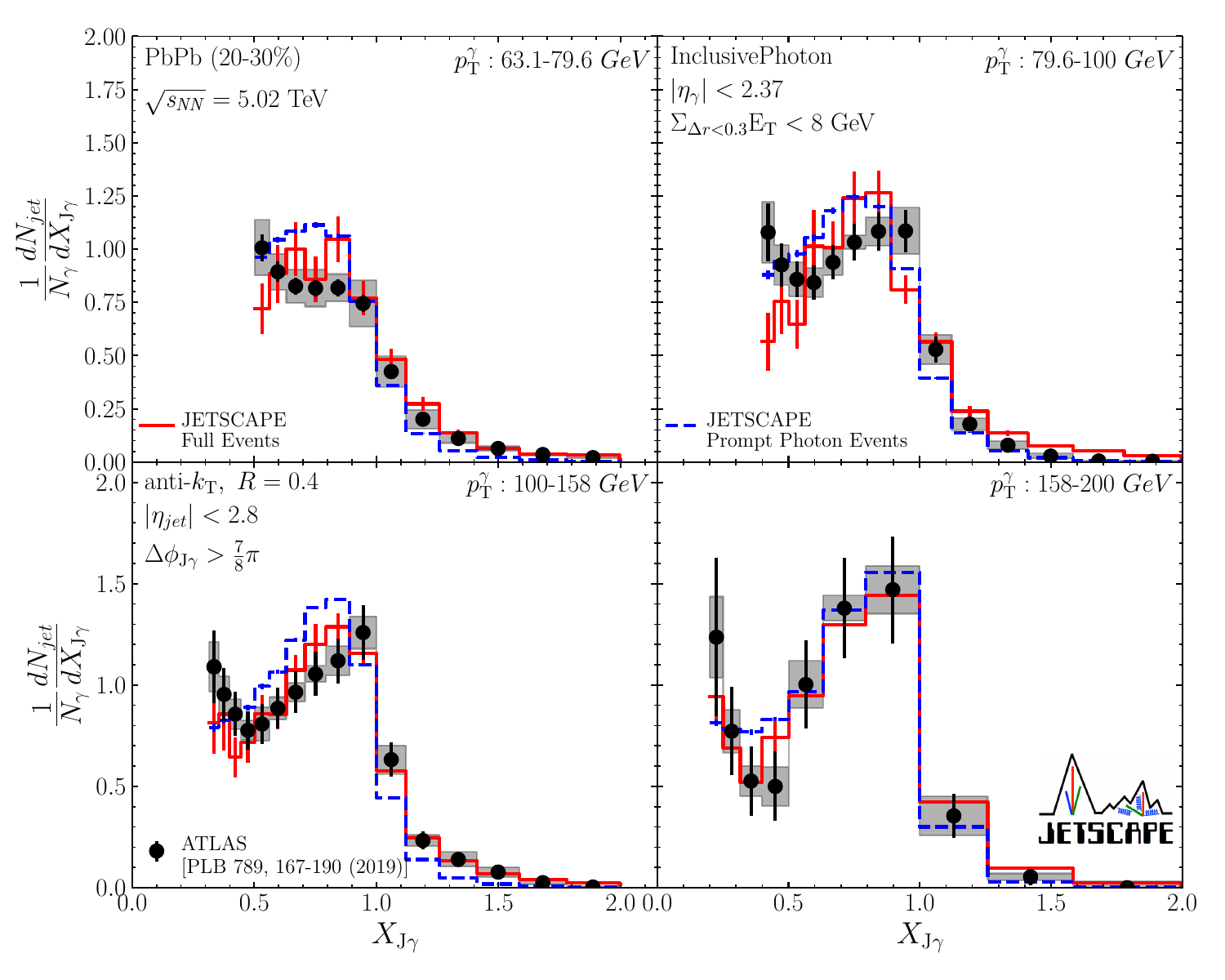}  
\caption{Same as Fig.~\ref{fig:Xj_ATLAS_0_10} for the $20\%$--$30\%$ centrality}
\label{fig:Xj_ATLAS_20_30}
\end{figure*}
Distributions of $X_{J\gamma}$ in Pb-Pb collisions at $\sqrt{s_{NN}}=5.02$~TeV are compared with CMS data~\cite{CMS:2017ehl} for the $0\%$--$30\%$ centrality in Fig.~\ref{fig:Xj_CMS_PbPb_0_30}, and with ATLAS data~\cite{ATLAS:2018dgb} for $0\%$--$10\%$ in Fig.~\ref{fig:Xj_ATLAS_0_10}, for $10\%$--$20\%$ in Fig.~\ref{fig:Xj_ATLAS_10_20}, and for $20\%$--$30\%$ in Fig.~\ref{fig:Xj_ATLAS_20_30}, respectively. 
The isolated photons are selected using the same kinematic cuts as for the $p$-$p$ collisions for comparison with the CMS results. 
For comparison with the ATLAS results, a transverse energy cut of $E_T^{\mathrm{iso}} < 8$ GeV inside a cone of $R_{\mathrm{iso}}=0.3$ around the photon is imposed while the rapidity cut remains the same as in the $p$-$p$ case.

The distributions in $p$-$p$ collisions 
exhibit a peak around or slightly below $X_{J\gamma} = 1$, approximating the kinematic limit defined by the primary contribution from prompt photon events. 
In comparison with CMS data (Fig.~\ref{fig:Xj_CMS_pp_0_30}), 
a relatively broad peak below $X_{J\gamma} = 1$ is observed, which can be attributed to smearing effects. This broadening occurs consistently in both distributions of full events and prompt photon events. 
On the other hand, comparisons with ATLAS unfolded results (Fig.~\ref{fig:Xj_ATLAS_pp}), in which the smearing is not applied, reveal a distinct peak near $X_{J\gamma} = 1$. This peak is prominent in both the full event and prompt photon event distributions, further indicating that prompt photon events are the dominant contributors to the peak position. 

In Pb-Pb collisions with medium effects, jet energy loss causes the peak position to shift toward lower values, becoming smeared or accumulating in the region of smaller $X_{J\gamma}$. The extent of this shift in the case of the full events is somewhat closer to the data, when the entire data set is considered, than in the case of just prompt photons. This once again highlights the importance of bremsstrahlung photons in photon-triggered jet events.

Across all cases, both in $p$-$p$ and Pb-Pb collisions, it is relatively clear that the values from the full event set are larger in the region where $X_{J\gamma}\gtrapprox 1$. 
In prompt photon productions, the maximum transverse momentum of the paired jet is approximately constrained by the transverse momentum of the photon. Consequently, in regions where the jet's transverse momentum exceeds that of the photon ($X_{J\gamma}\gtrapprox 1$), the prompt photon contribution becomes particularly small. 
This allows the contribution from non-prompt photons---despite being strongly suppressed by the back-to-back restriction with the relative azimuth angle cut $\Delta \phi_{J\gamma} > 7 \pi / 8$---to become relatively prominent.
Through normalization, this difference is reflected in relatively sizable discrepancies between the two event sets, also at small $X_{J\gamma}\lessapprox 1$, in some $p^{\gamma}_{T}$ regions both in $p$-$p$ and Pb-Pb collisions.

\subsection{Photon-jet azimuth correlation}
\label{Subsection:Correlation}

\begin{figure*}[htb!]
\centering
\includegraphics[width=0.99\textwidth]{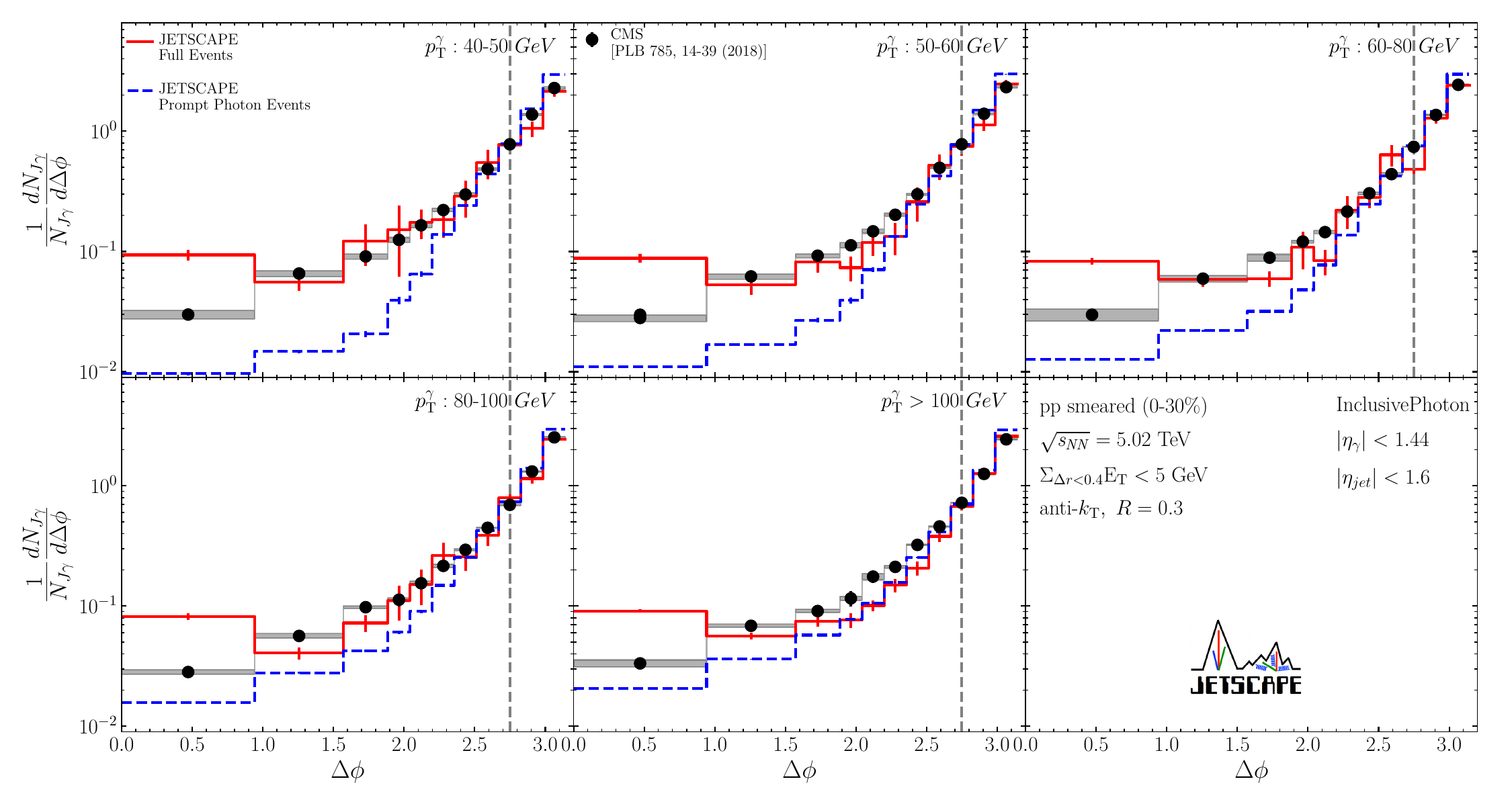}  
\caption{Photon-jet azimuth correlation for $p$-$p$ collisions at $\sqrt{s_{NN} }= 5.02$~TeV, smeared for $0\%$--$30\%$, for five different $p_T^{\gamma}$ intervals. 
The results from \jetscape\ for full events (solid lines) and prompt photon events (dashed lines) are compared with CMS data~\cite{CMS:2017ehl}. See text for details. }
\label{fig:dPhi_CMS_pp_0_30}
\end{figure*}
\begin{figure*}[htb!]
\centering
\includegraphics[width=0.99\textwidth]{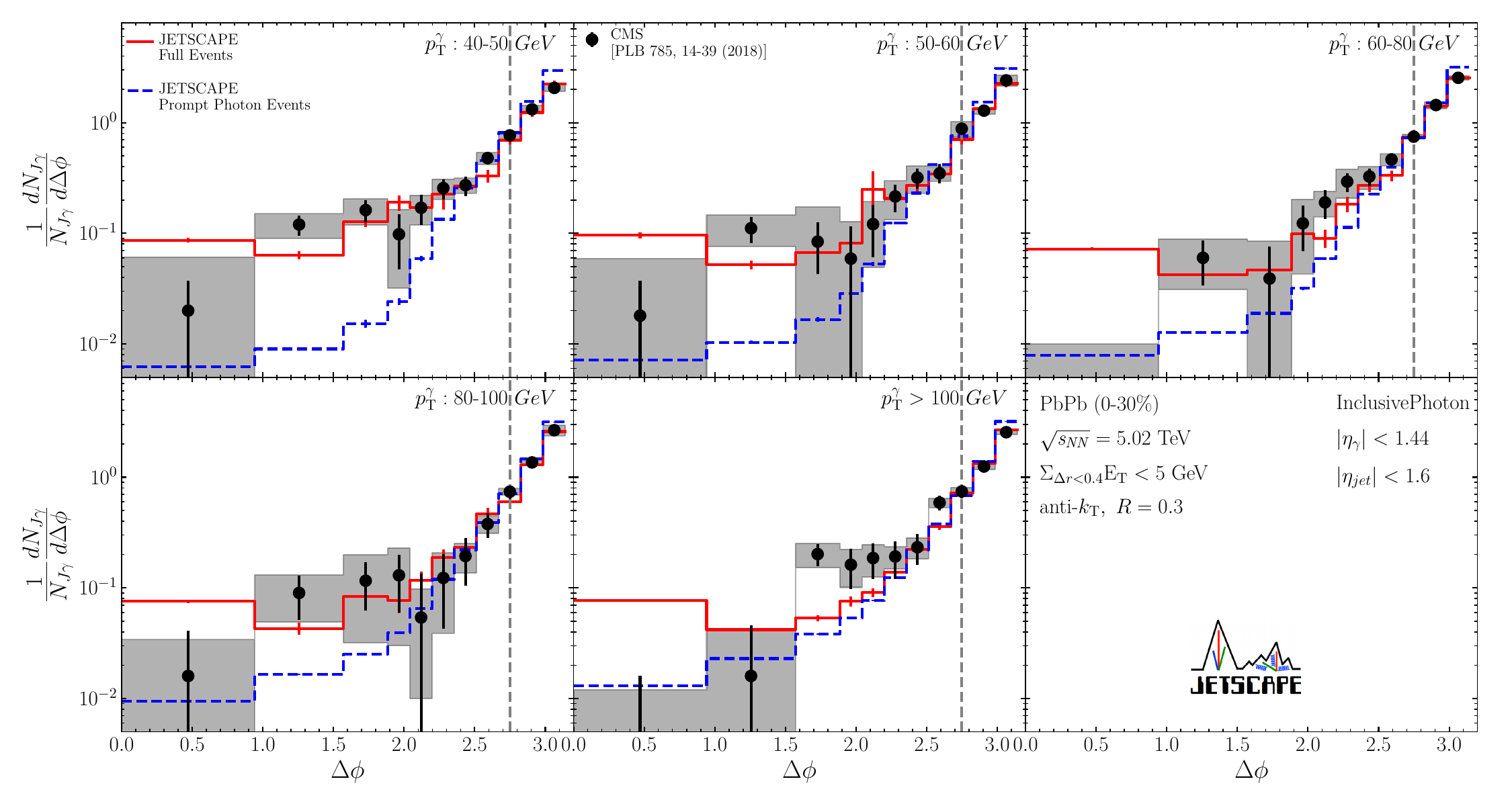}  
\caption{Photon-jet azimuthal correlation in $0\%$--$30\%$ Pb-Pb collisions at $\sqrt{s_{NN} }= 5.02$~TeV for five different $p_T^{\gamma}$ intervals. 
The results from \matter$+$\lbt\ within \jetscape\ for full events (solid lines) and prompt photon events (dashed lines) are compared with CMS data~\cite{CMS:2017ehl}. See text for details. }
\label{fig:dPhi_CMS_PbPb_0_30}
\end{figure*}
In this subsection, we present the results for the photon triggered jet azimuthal correlation: 
\begin{align}
\frac{1}{N_{\gamma}}\frac{dN_{\mathrm{jet}}}{d\Delta \phi}.
\label{eq:dphi_dist}
\end{align}
Here, the relative azimuth angle cut is not imposed. 
Figures~\ref{fig:dPhi_CMS_pp_0_30} and ~\ref{fig:dPhi_CMS_PbPb_0_30} show the $\Delta \phi$ distribution in $p$-$p$ and $0\%$--$30\%$ Pb-Pb collisions at $\sqrt{s_{NN}}=5.02$ TeV, respectively. 
Results are compared with CMS data~\cite{CMS:2017ehl}.

When a photon is produced from the initial hard scattering, the associated jet typically produces partons with a larger azimuthal separation. Since a photon does not interact with the medium, it is extremely rare for partons to appear on the nearside of the photon. As a result, significant suppression in the small $\Delta \phi$ region is commonly observed compared to experimental results, which use isolated photons, when only prompt-photon events are considered.

However, when total inclusive events with isolated photons are considered, better agreement with experimental results is expected. This trend is evident in Figures ~\ref{fig:dPhi_CMS_pp_0_30} and ~\ref{fig:dPhi_CMS_PbPb_0_30}, where prompt photon events exhibit significant suppression in the small $\Delta \phi$ region across all $p_T^{\gamma}$ intervals, as anticipated, while demonstrating better agreement in the large $\Delta \phi$ region.

 On the other hand, full events demonstrate excellent agreement with experimental results for all $p_T^{\gamma}$ intervals despite slightly larger statistical errors. The noticeable enhancement in the smallest $\Delta \phi$ bin might be due to the accumulation of soft particles in isolated non-prompt photons.

%%%%%% SUMMARY %%%%%%%%%%%%%
\section{Summary}
\label{Section:Summary}
In this paper, we have studied photon triggered jets in $p$-$p$ and Pb-Pb collisions at $\sqrt{s_{NN}}=5.02$ TeV by performing simulations within the \jetscape\ framework. 
For the $p$-$p$ case, the virtuality-ordered jet evolution with parton and photon radiation was simulated by the \matter\ module with medium effects turned off. 
For the Pb-Pb case, we employ the multistage setup of \matter$+$\lbt, where medium effects on parton radiation in the \matter\ module are activated, and the \lbt\ module handles transport evolution of partons at low virtuality, with the parameter set obtained from the previous study~\cite{JETSCAPE:2022jer}.

To investigate the contributions of non-prompt photons---predominantly bremsstrahlung photons, which arise from processes such as photon radiation in high virtuality showers---and decay photons to the yield of photon triggered jets, analyses were conducted using two different event sets. The first was the Full Event Set, which inclusively generates hard-scattering events encompassing all hard QCD channels. The second was the Prompt Photon Event Set, which contains only hard scattering events that produce prompt photons. 
In the full event set, photons arise from the prompt hard scattering process, from bremsstrahlung in the high virtuality \matter\ stage,  and in very rare cases from the decay of a hadron. Most decay photons do not pass the isolation test, and no bremsstrahlung photons are included in the lower virtuality \lbt\ stage. 

As it is extremely rare to observe an isolated photon in an inclusive event, having a sufficiently large number of events is important to reduce statistical uncertainty. However, for Prompt Photon Events, the overwhelming majority of events involve isolated photons, requiring only a relatively small number of total events that need to be simulated. Throughout this study, it becomes clear that full events capture most of the features of the experimental results, though with noticeably (and unavoidably) larger uncertainty compared to the Prompt Photon Events.

In the $R_{AA}$ for photon triggered jets, where a back-to-back azimuthal angular constraint was applied between the photon and the jet, one observes a slight improvement in the description of the data using prompt and non-prompt photons (full events) compared to just prompt photon triggers. Both event sets yielded values comparable to experimental results with the parameter set tuned with the single high-$p_{T}$ particle and jet $R_{AA}$s in Ref.~\cite{JETSCAPE:2022jer}, with the full events showing an overall improvement compared to prompt photon events. 
This was also evident in the medium modification of the $X_{J\gamma}$ distribution, where a back-to-back constraint was applied, as both event sets exhibited similar shifts and distortions of the peaks, with the full events displaying a somewhat better fit to the data than the case of prompt photons.

Due to kinematic constraints, prompt photon-jet pairs produced in hard scattering are significantly suppressed in the region $X_{J\gamma} > 1$. In this region, despite the back-to-back constraint, contributions from non-prompt photons increase the yields of photon triggered jets in both $p$-$p$ and Pb-Pb collisions. The effects of non-prompt photons not only provide a more accurate description of the $X_{J\gamma} > 1$ region but also have a slight impact on the overall distribution through normalization. 

The photon-jet azimuthal correlation, without imposing a back-to-back azimuthal constraint, clearly demonstrates a significant contribution from non-prompt photons. In particular, in the same azimuthal hemisphere as the jet ($\Delta \phi > \pi$), jet-photon pairs entirely dominated by non-prompt photons prevail. Including this contribution is essential for describing the experimental results. Furthermore, this effect, through normalization, influences the distribution across the entire angular range.

The improvement of comparisons with experimental data for the compute intensive case of full events including prompt and bremsstrahlung photons (using the same parameter set as in Refs.~\cite{JETSCAPE:2022jer,JETSCAPE:2022hcb,JETSCAPE:2023hqn}) makes the observables included in this paper less than ideal for inclusion in the next set of Bayesian analysis. Their calculation involves simulating many more events, and they will likely not introduce any new tension in the comparison between experimental data and the model of Ref.~\cite{JETSCAPE:2024ofm}. In an upcoming companion paper we will demonstrate that photon triggered jet substructure observables may indeed provide a data set that under certain conditions can be reliably simulated with only prompt photon events, thus making $\gamma$-triggered jet substructure more ideal for inclusion in a more extensive Bayesian analysis. 

While not ideal for a Bayesian analysis, the successful comparison between our simulations and the experimental data, over a range of jet energy, centrality, and energy of collision strongly support the multistage model of Ref.~\cite{JETSCAPE:2022jer,JETSCAPE:2022hcb,JETSCAPE:2023hqn}. All comparisons to experimental data in this paper are parameter-free postdictions. No parameter has been tuned to any data point. This adds strong confirmation for the efficacy of the multistage model of energy loss.

%%%%%% ACKNOWLEDGEMENT %%%%%
\section*{ACKNOWLEDGMENTS}

%thank specific people who helped:
The authors thank Nihar Sahoo for providing experimental results for the photon-triggered jet $I_{AA}$. 

%thank funding agencies:
This work was supported in part by the National Science Foundation (NSF) within the framework of the JETSCAPE collaboration, under grant number OAC-2004571 (CSSI:X-SCAPE). It was also supported under   \rm{PHY-1516590}, \rm{PHY-1812431} and \rm{PHY-2413003} (R.J.F., M.Ko., C.P. and A.S.); it was supported in part by the US Department of Energy, Office of Science, Office of Nuclear Physics under grant numbers \rm{DE-AC02-05CH11231} (X.-N.W. and W.Z.), \rm{DE-AC52-07NA27344} (A.A., R.A.S.), \rm{DE-SC0013460} (A.K., A.M., C.Sh., I.S., C.Si and R.D.), \rm{DE-SC0021969} (C.Sh. and W.Z.), \rm{DE-SC0024232} (C.Sh. and H.R.), \rm{DE-SC0012704} (B.S.), \rm{DE-FG02-92ER40713} (J.H.P. and M.Ke.), \rm{DE-FG02-05ER41367} (C.Si, D.S. and S.A.B.), \rm{DE-SC0024660} (R.K.E), \rm{DE-SC0024347} (J.-F.P. and M.S.). The work was also supported in part by the National Science Foundation of China (NSFC) under grant numbers 11935007, 11861131009 and 11890714 (Y.H. and X.-N.W.), by the Natural Sciences and Engineering Research Council of Canada (C.G., S.J., and G.V.),  by the University of Regina President's Tri-Agency Grant Support Program (G.V.), by the Canada Research Chair program (G.V. and A.K.) reference number CRC-2022-00146, by the Office of the Vice President for Research (OVPR) at Wayne State University (Y.T.), by JSPS KAKENHI Grant No.~22K14041 (Y.T.), and by the S\~{a}o Paulo Research Foundation (FAPESP) under projects 2016/24029-6, 2017/05685-2 and 2018/24720-6 (M.L.). C.Sh., J.-F.P. and R.K.E. acknowledge a DOE Office of Science Early Career Award. I.~S. was funded as part of the European Research Council project ERC-2018-ADG-835105 YoctoLHC, and as a part of the Center of Excellence in Quark Matter of the Academy of Finland (project 346325).

%thank computational facilities:
Calculations for this work used the Wayne State Grid, generously supported by the Office of the Vice President of Research (OVPR) at Wayne State University.
The bulk medium simulations were performed using resources provided by the Open Science Grid (OSG) \cite{Pordes:2007zzb, Sfiligoi:2009cct}, which is supported by the National Science Foundation award \#2030508.
Data storage was provided in part by the OSIRIS project supported by the National Science Foundation under grant number OAC-1541335.

%%%%%% APPENDIX %%%%%%%%%%%%
% \hypertarget{Appen}{}
% \appendix*

%%%%%% REFERENCES %%%%%%%%%%
\bibliography{main,manual,misc}
\end{document}